\documentclass[letterpaper, 10 pt, conference]{ieeeconf}  

\IEEEoverridecommandlockouts                              

\usepackage{graphics} 
\usepackage{epsfig} 
\usepackage{mathptmx} 
\usepackage{times} 
\usepackage{amsmath} 
\usepackage{amssymb}  
\usepackage{url}
\usepackage{array}
\usepackage{algorithmic}
\usepackage{algorithm}
\usepackage{subcaption}
\usepackage{cite}
\usepackage{bm}
\mathchardef\mhyphen="2D
\newcommand{\PropMethod}{{X-Selector}}

\usepackage[table,xcdraw]{xcolor}

\title{\LARGE \bf
  User Decision Guidance with Selective Explanation Presentation \\ from Explainable-AI
}

\author{Yosuke Fukuchi$^{1}$ and Seiji Yamada$^{2,3}$
\thanks{$^{1}$Faculty of Systems Design, Tokyo Metropolitan University, Tokyo, Japan
        {\tt\small fukuchi@tmu.ac.jp}.
$^{2}$Digital Content and Media Sciences Research Division, National Institute of Infomatics, Tokyo, Japan.
$^{3}$The Graduate University for Advanced Studies, SOKENDAI, Tokyo, Japan.
This work was supported in part by JST CREST Grant Number JPMJCR21D4 and JSPS KAKENHI Grant Number JP24K20846.}%
}

\begin{document}

\maketitle

\thispagestyle{empty}
\pagestyle{empty}


\begin{abstract}
This paper addresses the challenge of selecting explanations for XAI (Explainable AI)-based Intelligent Decision Support Systems (IDSSs). 
IDSSs have shown promise in improving user decisions through XAI-generated explanations along with AI predictions,
and the development of XAI made it possible to generate a variety of such explanations.
However, how IDSSs should select explanations to enhance user decision-making remains an open question.
This paper proposes {\PropMethod}, a method for selectively presenting XAI explanations.
It enables IDSSs to strategically guide users to an AI-suggested decision
by predicting the impact of different combinations of explanations on a user's decision and selecting the combination that is expected to minimize the discrepancy between an AI suggestion and a user decision.
We compared the efficacy of {\PropMethod} with two naive strategies (all possible explanations and explanations only for the most likely prediction) and two baselines (no explanation and no AI support).
The results suggest the potential of {\PropMethod} to guide users to AI-suggested decisions and improve task performance under the condition of a high AI accuracy.
\end{abstract}

\section{Introduction}
Intelligent Decision Support Systems (IDSSs)~\cite{idss}, empowered by Artificial Intelligence (AI), have the potential to help users make better decisions by introducing explainability into their support.
An increasing number of methods have been proposed for achieving explainable AIs (XAIs)~\cite{8466590},
and the development of large language models (LLMs) has also made it possible to generate various post-hoc explanations that justify AI predictions.
Previous studies have integrated such XAI methods into IDSSs and shown their effectiveness in presenting explanations along with AI predictions in diverse applications~\cite{10.1145/3610218,electronics12214430}.

Now that IDSSs can have a variety of explanation candidates, a new question arises as to which explanations an IDSS should provide in dynamic interaction.
Explanation is a complex cognitive process~\cite{MILLER20191}.
Although XAI explanations can potentially guide users to make better decisions, there is also a risk of having negative effects on explainees' decisions.
Various causes including explanation uninterpretability~\cite{maehigashi2023modeling,maehigashi2023roman}, 
information overload~\cite{ferguson2022explanations,herm2023impact}, and contextual inaccuracy~\cite{10.1145/3301275.3302316}
can affect users and thus the performance of decision-making.
A subtle difference in the nuance of a linguistic explanation can also have a different impact and sometimes mislead user decisions when influenced by the context, the status of the task, and the cognitive and psychological status of the users.
Conversely, we can expect that IDSSs can be greatly enhanced if they can strategically select explanations that are likely to lead users to better decisions while taking the situation into account.

To address the question of how IDSSs can select explanations to improve user decisions, 
this paper proposes {\PropMethod}, a method for dynamically selecting which explanations to provide along with an AI prediction.
The main characteristic of {\PropMethod} is that it predicts how explanations affect user decision-making for each trial and attempts to guide users to an AI-suggested decision referring to the prediction results.
The design of guiding users with explanation selection is inspired by libertarian paternalism~\cite{b35e72fa-fff9-37d3-a508-45875042aa96},
the idea of influencing one's choices to better ones while embracing the autonomy of their decision-making.

This paper also reports user experiments that simulated stock trading with an XAI-based IDSS.
In a preliminary experiment, we compared two naive but common strategies---ALL (providing all possible explanations) and ARGMAX (providing only explanations for the AI's most probable prediction)---against baseline scenarios 
providing no explanations or no decision support.
The results suggest that the ARGMAX strategy works better with high AI accuracy, and ALL is more effective when AI accuracy is lower,
indicating that the strategy for selecting explanations affects user performance.
In the second experiment, we compared the results of explanations selected by {\PropMethod} with ARGMAX and ALL.
The results indicate the potential of {\PropMethod}'s selective explanations to more strongly lead users to suggested decisions and to outperform ARGMAX when AI accuracy is high.


\section{Background}
\subsection{XAIs for deep learning models}\label{ss:XAI}
While various methods such as Fuzzy Logic and Evolutionary Computing have been introduced to IDSSs, this paper targets IDSSs with Deep Learning (DL) models.
IDSSs driven by DL models are capable of dealing with high-dimensional data such as visual images
and are actively studied in diverse fields \cite{KRAUS201738,9280654,10.1371/journal.pone.0213007}.
Due to their blackbox nature, explainability for DL models is also an area of active research, 
and this can potentially offer benefits for IDSSs.

There are various forms of explanations depending on the nature of the target AI.
Common explanations for visual information processing AIs include presenting saliency maps. 
The class activation map (CAM) is a widely used method for visualizing a saliency map of convolutional neural network (CNN) layers~\cite{Zhou_2016_CVPR}.
It identifies the regions of an input image that contribute the most for a model to classify the image into a particular class.

Language is also a common modality of XAIs, and free-text explanation is becoming rapidly available thanks to the advance of LLMs~\cite{wiegreffe-etal-2022-reframing}.
LLMs can generate \textit{post-hoc} explanations for AI predictions.
Here, post-hoc means that the explanations are generated after the AI's decision-making process has occurred, as opposed to \textit{intrinsic} methods that generate explanations in an integral part of that process~\cite{8466590}.

\subsection{Human-XAI interaction}
The theme of this study involves how to facilitate users' appropriate use of AI.
Avoiding human over/under-reliance on an AI is a fundamental problem of human-AI interaction~\cite{doi:10.1518/001872097778543886}.
Here, over-reliance is a state in which a human overestimates the capability of an AI and blindly follows its decision, 
whereas under-reliance is a state in which a human misuses an AI even though it can perform well.

Although explanation is believed to generally help people appropriately use AI by providing transparency in AI predictions,
previous studies suggest that XAI explanations do not always work positively~\cite{10.1145/3579605}.
Maehigashi et al. demonstrated that presenting AI saliency maps has different effects on user trust in an AI depending on the task difficulty and the interpretability of the saliency map~\cite{maehigashi2023modeling}.
Herm revealed that the type of XAI explanation strongly influences users' cognitive load, task performance, and task time~\cite{herm2023impact}.
Panigutti et al. conducted a user study with an ontology-based clinical IDSS,
and they found that the users were influenced by the explanations despite the low perceived explanation quality~\cite{10.1145/3491102.3502104}.
These results suggest potential risks of triggering under-reliance with explanations or, conversely, leading to users blindly following explanations from an IDSS even if the conclusion drawn from the explanations is incorrect.

This study aims to computationally predict how explanations affect user decisions
in order to avoid misleading users and encourage them to make better decisions by selecting explanations.
Work by Wiegreffe et al.~\cite{wiegreffe-etal-2022-reframing} shares a similar concept with this study.
They propose a method of evaluating explanations generated by LLMs by predicting human ratings of their acceptability.
This approach is pivotal in understanding how users perceive AI-generated explanations.
However, our study diverges by focusing on the behavioral impacts of these explanations on human decision-making.
We are particularly interested in how these explanations can alter the decisions made by users getting an IDSS's support, rather than just their perceptions of the explanations.
Another relevant study, Pred-RC~\cite{fukuchi2023selectively,10343151}, aims to predict the effect of explanations of AI performance so that users can avoid over/under-reliance.
It dynamically predicts a user's binary judgment of whether s/he assigns a task to the AI and selects explanations that guide him/her to better assignment.
{\PropMethod} aims to take a further step to predict concrete decisions taking the effects of explanations into account and proactively influences them to improve the performance of each decision-making.

\section{\PropMethod}
\subsection{Overview}
This paper proposes {\PropMethod}, a method for dynamically selecting explanations for AI predictions.
The idea of {\PropMethod} is that it predicts user decisions under possible combinations of explanations and chooses the best one that 
is predicted to minimize the discrepancy between the decision that the user is likely to make and the AI-suggested one.

\subsection{Algorithm}
The main components of {\PropMethod} are UserModel and $\pi$.
UserModel is a model of a user who makes decision $d_u$.
{\PropMethod} uses it for user decision prediction:
\begin{equation}
  \mathrm{UserModel}(\bm{c}, \bm{x}, d) = P(d_u = d| \bm{c}, \bm{x}).
\end{equation}
The output of UserModel is represented as a probaility distribution of $d_u$ conditioned by $\bm{c}$ and $\bm{x}$,
where $\bm{x} \in \bm{X}$ is a combination of explanations to be presented to the user, and $\bm{c}$ represents all the other contextual information including AI predictions, task status, and user status. 
In this paper, we developed a dataset of ($\bm{c}, \bm{x},$ and $d_u$) and prepared a machine learning model that was trained with the dataset for implementing this.

In addition, {\PropMethod} has a policy $\pi$, which considers a decision $d_\mathrm{AI}$ based on $\bm{c}$.
This inference is done in parallel with user decision-making:
\begin{equation}
  \pi(\bm{c}, d) = P(d_\mathrm{AI} = d| \bm{c}).
\end{equation}

{\PropMethod} aims to minimize the discrepancy between $d_u$ and $d_\mathrm{AI}$ by comparing the effect of each $\bm{x}$ on $d_u$.
The selected combination $\hat{\bm{x}}$ is calculated as:
\begin{equation}\label{eqn:loss_func}
   \hat{\bm{x}} = \mathrm{argmin}_{\bm{x}} | E_{d \sim {\mathrm{UserModel}}(\bm{c}, \bm{x}, d)}[d] - E_{d \sim \pi(\bm{c}, d)}[d]|.
\end{equation}
To calculate this equation, {\PropMethod} simulates how $\bm{x}$ will change $d_u$ using UserModel and aims to choose the best one that guides $d_u$ to $d_{\mathrm{AI}}$ the most.

\subsection{Implementation}
\subsubsection{Task}
\begin{figure}[t]
  \begin{center}
    \begin{minipage}[t]{0.9\linewidth}
      \centering
      \includegraphics[width=0.6\linewidth]{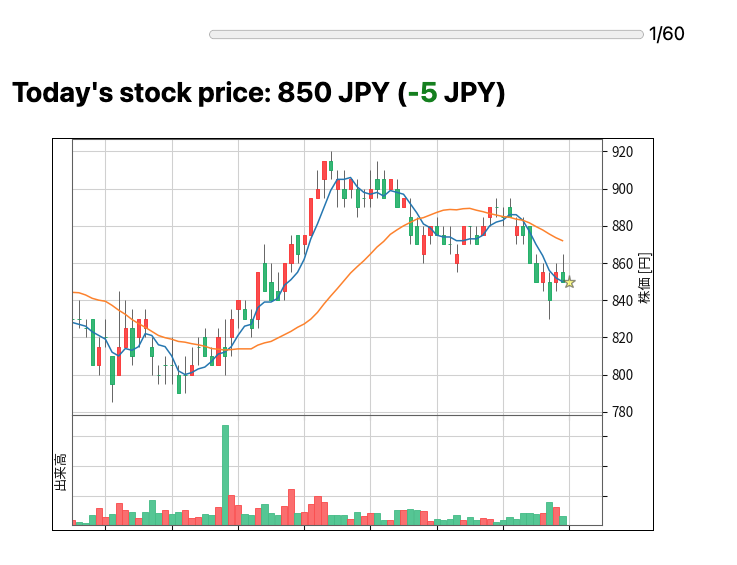}
      \subcaption{Chart}
      \label{fig:screenshotA}
    \end{minipage}
    \begin{minipage}[t]{0.9\linewidth}
      \centering
      \includegraphics[width=0.8\linewidth]{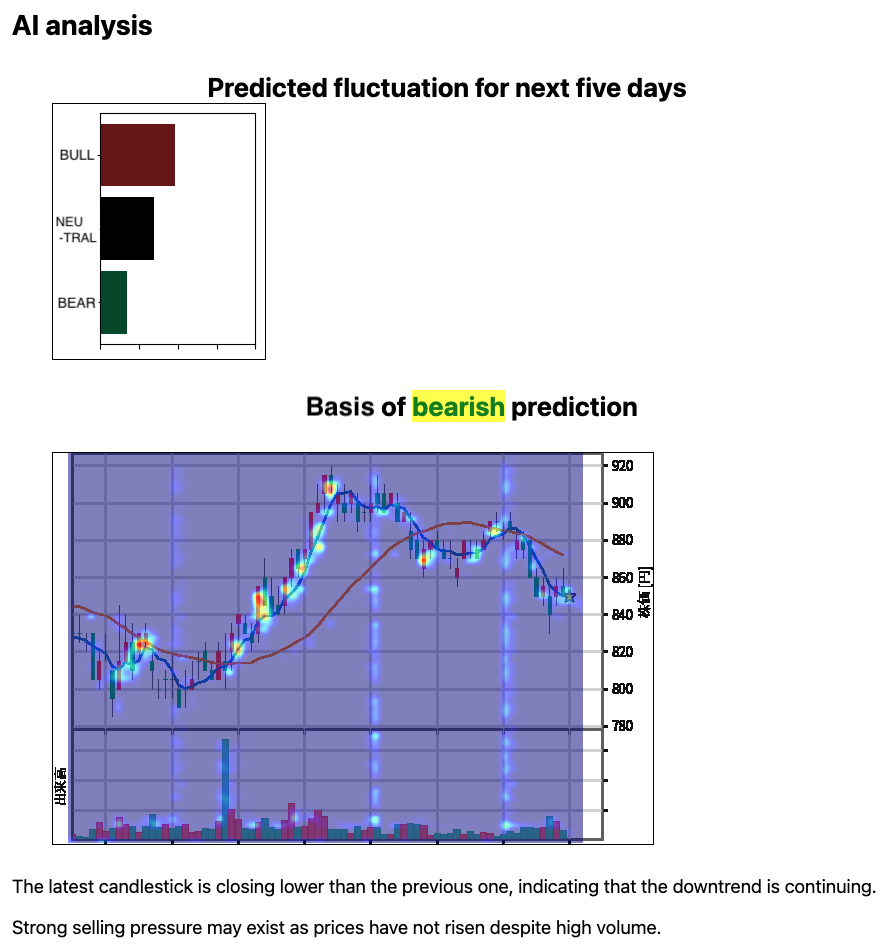}
      \subcaption{Examples of StockAI's prediction and its explanations}
      \label{fig:screenshotB}
    \end{minipage}
    \caption{Screenshots of trading simulator}\label{fig:screenshots}
      \vspace{-12pt}
  \end{center}
\end{figure}

We implemented {\PropMethod} in a stock trading simulator in which users get support from an XAI-based IDSS.
Figure \ref{fig:screenshots} shows screenshots of the simulator.
The simulation was conducted on a website.
Participants were virtually given three million JPY and traded stocks for 60 days with a stock price chart, AI prediction of the future stock price, and explanations for the prediction.

In the simulation, participants checked the opening price and a price chart for each day and decided whether to buy stocks with the funds they had, sell stocks they had, or hold their position.
In accordance with Japan's general stock trading system, participants could trade stocks in units of 100 shares.
Participants were asked to show their decision twice a day to clarify the influence of the explanations.
They were first asked to decide an initial order $d'$, that is, the amount of trade only with chart information and without the support of the IDSS.
Then, the IDSS showed a bar graph that indicated the output of a stock price prediction model and its explanations.
We did not explicitly show $d_{\mathrm{AI}}$ to enhance the autonomy of users' decision-making,
which is inspired by libertarian paternalism~\cite{b35e72fa-fff9-37d3-a508-45875042aa96},
the idea of affecting behavior while respecting freedom of choice as well.
However, we can easily extend {\PropMethod} to a setting in which $d_{\mathrm{AI}}$ is given to users by including it with $\bm{c}$ (when you always show $d_{\mathrm{AI}}$) or $\bm{x}$ (when you want to selectively show $d_{\mathrm{AI}}$). 
Finally, they input their final order $d$.
After this, the simulator immediately transited to the next day.
The positions carried over from the final day were converted into cash on the basis of the average stock price over the next five days to calculate the participants' total performance.

\subsubsection{StockAI}
In the task, an IDSS provides a prediction of a stock price prediction model (StockAI) as user support.
StockAI is a machine-learning model that is designed to predict the average stock price in the next five business days,
and we used its prediction as the target of the explanation provided to users.
StockAI predicts future stock prices on the basis of an image of a candlestick chart.
Although using a candlestick chart as an input does not necessarily lead to better performance than modern approaches proposed in the latest studies~\cite{7979885},
we chose this because of the better understandability of saliency maps generated with the model as an explanation of AI predictions.
Note that the aim of this research is not building a high-performance prediction model but investigating the interaction between a human and an AI whose performance is not necessarily perfect.

For the implementation of StockAI, we used ResNet-18~\cite{he2016deep}, a deep-learning visual information processing model, using the PyTorch library~\footnote{https://pytorch.org/}.
The StockAI is trained in a supervised manner;
it classifies the ratio of the future stock price to the opening price of the day into three classes: BULL (over +2\%), NEUTRAL (from -2 to +2\%), and BEAR (under -2\%).
The prediction results are presented as a bar graph of the probability distribution for each class, which hereafter denoted as $p$.
For the training, we collected the historical stock data (from 2018/5/18 to 2023/5/16) of companies that are included in the Japanese stock index Nikkei225.
We split the data by stock code, with three-quarters of the data as the training dataset and the remainder as the test.
The accuracy with which the model was able to predict the correct class among the three classes was 0.474,
and the accuracy for binary classification, or the matching rate of the sign of the expected value of the model's prediction and that of actual fluctuations, was 0.63
for the test dataset.

\subsubsection{Explanations}
We prepared two types of explanations: saliency maps and free-texts.
We applied CAM-based methods available in the pytorch-grad-cam package~\footnote{https://github.com/frgfm/torch-cam} to StockAI 
and adopted Score-CAM~\cite{Wang_2020_CVPR_Workshops} because it most clearly visualizes saliency maps of StockAI.
Because CAM-based methods can generate a saliency map for each prediction class, three maps were acquired for each prediction.
Let $\bm{x}_\mathrm{map}$ be the set of the acquired maps.

In addition, we created a set of free-text explanations based on the GPT-4V model in the Open-AI API~\cite{openai2023gpt4}, which allows images as input.
We input a chart with a prompt that asked GPT-4V to generate two explanation sentences that justify each prediction class (BULL, NEUTRAL, and BEAR).
Therefore, we acquired six sentences in total for each chart.
Let us denote the set of them by $\bm{x}_\mathrm{text}$.

As a result, three saliency maps and six free-text explanations were available for each trading day,
and {\PropMethod} considered $2^9 = 512$ combinations of the selected explanations ($\hat{\bm{x}} \subseteq \bm{x}_\mathrm{map} \cup \bm{x}_\mathrm{text}$).

\subsubsection{Models}
\begin{figure}
  \begin{center}
    \includegraphics[width=0.5\linewidth]{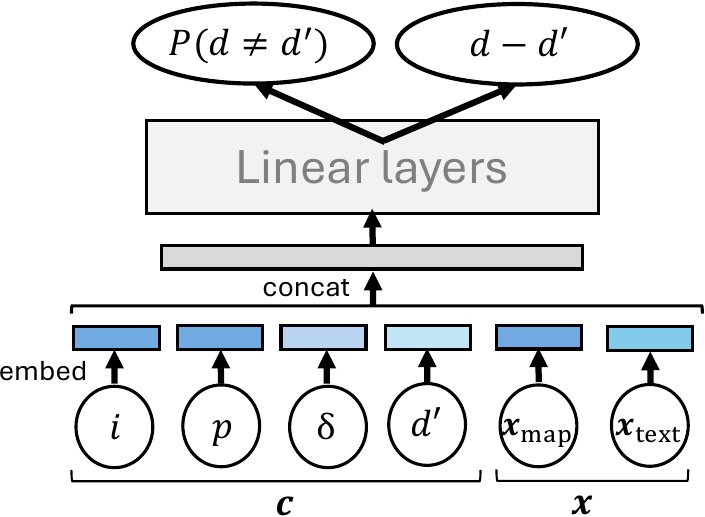}
    \caption{Structure of UserModel}\label{fig:user_model}
  \end{center}
      \vspace{-12pt}
\end{figure}
We implemented UserModel with a deep learning model (Figure~\ref{fig:user_model}).
The input of UserModel is a tuple $(\bm{c}, \bm{x})$.
$\bm{c}$ includes four variables: date $i$, StockAI's prediction $p$, total rate $\delta$, and initial order $d'$.
$i$ is a categorical variable that embeds the context of the day such as the stock price.
$p$ is a three-dimensional vector that corresponds to the values in the bar graph (Fig.~\ref{fig:screenshotB}).
$\delta$ is the percentage increase or decrease of the user's total asset from the initial amount.
$i$ and the other variables are encoded in 2048-dimensional vectors $(h_i, h_p, h_r, h_{d'})$ with the Embedding and Linear modules implemented in PyTorch, respectively.

Let us denote $\bm{x}_\mathrm{map}$ and $\bm{x}_\mathrm{text}$ as a set $\{(x, cls, flag)\}$, where $x$ is the raw data of an explanation,
and $cls \in \{\textrm{BULL}, \textrm{NEUTRAL}, \textrm{BEAR}\}$. 
$flag = 1$ if $x$ is to be presented, and $0$ when hidden.
$\bm{x}_\mathrm{map}$ and $\bm{x}_\mathrm{text}$ are also encoded in 2048-dimensional vectors $(h_{\mathrm{map}}, h_{\mathrm{text}})$:
\begin{align}
   h_{\mathrm{map}} &=  \sum_{\substack{x, cls, \\ flag}  \in \bm{x}_\mathrm{map}} flag \cdot (CNN(x) \odot ClsEmbedding(cls)),  \nonumber \\
   h_{\mathrm{text}} &= \sum_{\substack{x, cls, \\ flag} \in \bm{x}_\mathrm{text}} flag \cdot (TextEncoder(x) \odot ClsEmbedding(cls)), \nonumber
\end{align}
where $\odot$ denotes an element-wise product. $CNN$ is a three-layer CNN model.
For $TextEncoder$, we used the E5 (embeddings from bidirectional encoder representations) model~\cite{wang2022text} with pretrained parameters\footnote{https://huggingface.co/intfloat/multilingual-e5-large}.

All embedding vectors ($h_i, h_p, h_r, h_{d'}, h_{\textrm{map}}, h_\textrm{text}$) are concatenated and input to a three-layer linear model.
To extract the influence of explanations, the model was trained to predict not $d$ but the difference $d - d'$.
In our initial trial, UserModel always predicted $d - d'$ to be nearly 0 due to the distributional bias,
so we added an auxiliary task of predicting whether $d = d'$ and trained the model for predicting $d - d'$ only when $d \neq d'$.
The expected value of $d_u$ in Equation \ref{eqn:loss_func} is $P(d \neq d') \cdot (d - d') + d'$.

$\pi$ was acquired with the deep deterministic policy gradient, a deep reinforcement learning method~\cite{pmlr-v48-gu16}.
We simply trained $\pi$ to decide $d$ to maximize assets on the basis of $p$ for the training dataset.
The reward for the policy is calculated as the difference in total assets between the current day and the previous day.

\section{Experiments}
\subsection{Preliminary experiment}
\subsubsection{Procedure}
We conducted a preliminary experiment to investigate the performance of users who were provided explanations with two naive strategies (ALL and ARGMAX).
ALL shows all the nine explanations available for each day, and ARGMAX selects explanations for StockAI's most probable prediction.
To ensure the quality of the explanations, we also prepared two baselines:
ONLY\_PRED shows $p$ but does not provide any explanations. In PLAIN, participants received no support from the IDSS and acted on their own.

For simulation, we chose a Japanese general trading company (code: 2768) from the test dataset
on the basis of the common stock price range (1,000 - 3,000 JPY) and its high volatility compared with the other Nikkei225 companies.

Because we had anticipated that the accuracy of StockAI would affect the result, we prepared two scenarios: high-accuracy and low-accuracy.
We calculated the moving average of the accuracy of StockAI with a window size of 60 and chose two sections for them.
The accuracy of StockAI for high-accuracy was 0.750, which was the highest, and that for low-accuracy was 0.333, the chance level of three-class classification.

We recruited participants to join the simulation with compensation of 220 JPY through Lancers\footnote{https://lancers.jp/}, a Japanese crowdsourcing platform,
and got 336 participants.
The participants were first provided pertinent information, and 325 consented to the participation.
We gave them instructions on the task and gave basic explanations about stock charts and the price prediction AI.
We instructed the participants to increase the given three million JPY as much as possible by trading with the IDSS's support.
To motivate them, we told them that additional rewards would be given to the top performers.
We did not notify them of the amount of the additional rewards and the number of participants who got them.
We asked six questions to check their comprehension of the task.
34 participants who failed to answer correctly were excluded from the task.
After familiarization with the trading simulator, the participants traded for 60 virtual days successively.
242 participants completed the task (152 males, 88 females, and 2 did not answer; aged 14-77, $M = 42.8, SD = 10.1$).
Table \ref{table:sample-size} gives details on the sample sizes. 
\begin{table}[]
\caption{Sample sizes of preliminary experiment}\label{table:sample-size}
\begin{center}
\begin{tabular}{l|l|l|l|l}
                 & ALL & ARGMAX & ONLY\_PRED & PLAIN \\ \hline
High-accuracy &     39 &     40 &         34 &    41 \\ \hline
Low-accuracy  &     31 &     34 &         34 &    38 
\end{tabular}
\end{center}
\vspace{-12pt}
\end{table}

\subsubsection{Result}

\begin{figure}[t]
  \begin{center}
    \begin{minipage}[b]{0.75\linewidth}
      \centering
      \includegraphics[width=0.99\linewidth]{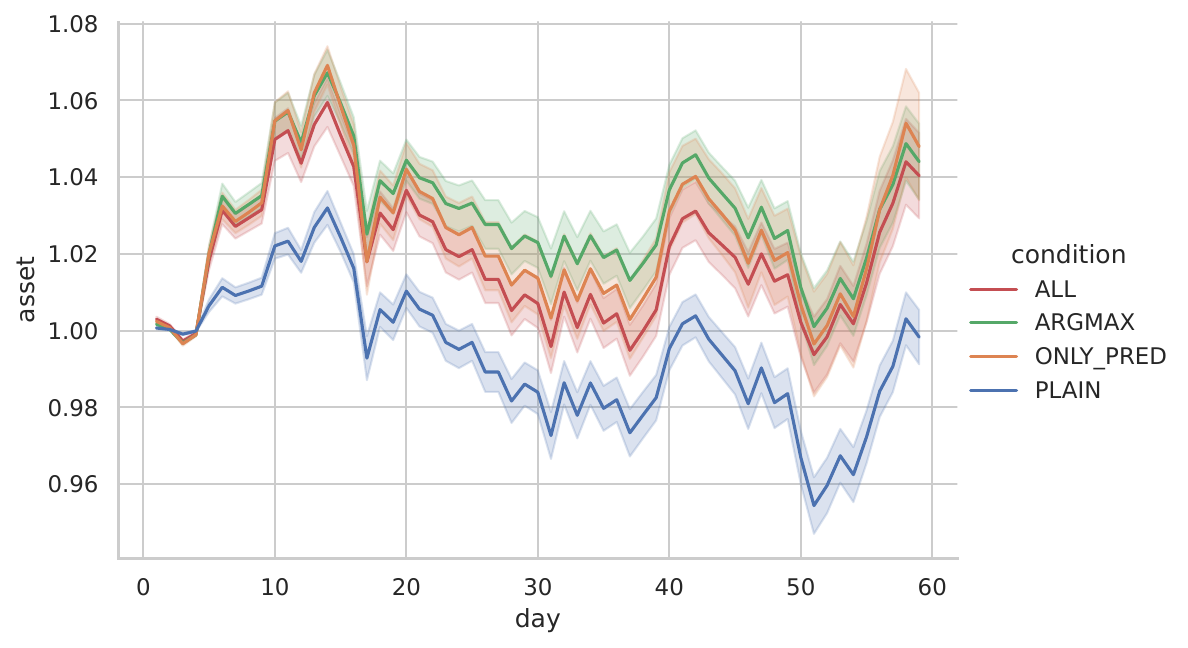}
      \subcaption{High-accuracy}
      \label{fig:4a}
    \end{minipage}
    \begin{minipage}[b]{0.75\linewidth}
      \centering
      \includegraphics[width=0.99\linewidth]{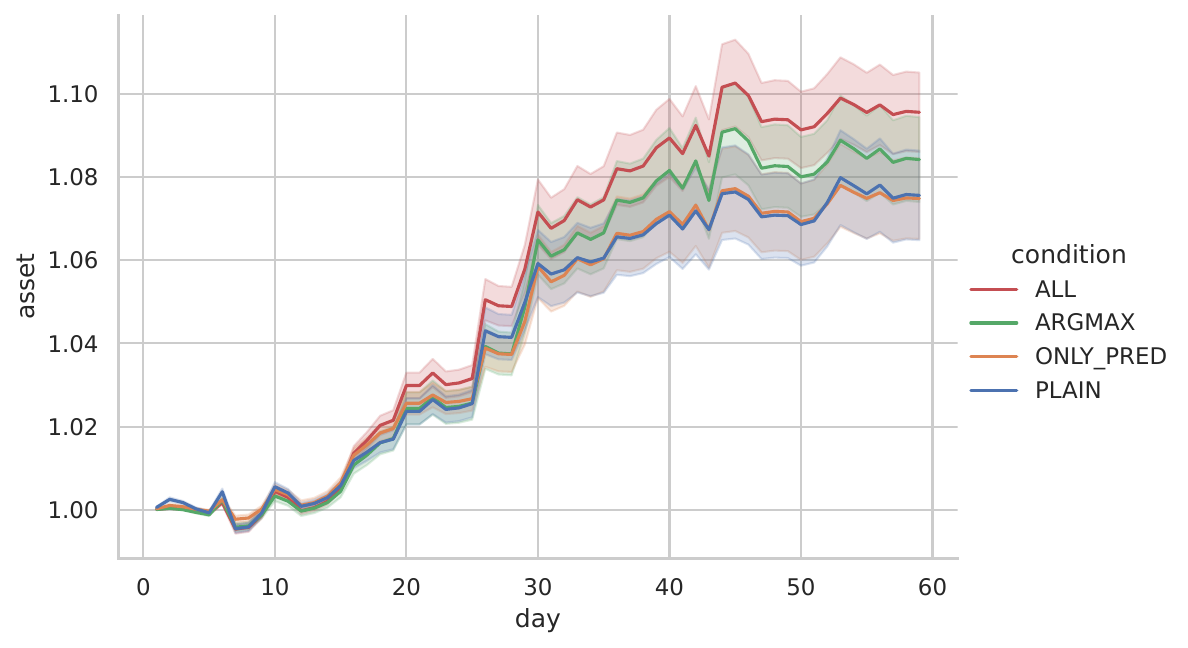}
      \subcaption{Low-accuracy}
      \label{fig:4b}
    \end{minipage}
    \caption{Comparisons of baseline total assets. Error bands represent standard errors.}\label{fig:preliminary}
      \vspace{-12pt}
  \end{center}
\end{figure}

Figure \ref{fig:preliminary} shows the changes in the participants' performance.
A conspicuous result is the underperformance of PLAIN, particularly in the high-accuracy scenario.
ONLY\_PRED performed well for high-accuracy, but could not outperform PLAIN for low-accuracy.
This suggests that presenting $p$ alone contributes to improving performance only when it has enough accuracy.

ALL and ARGMAX showed different results between the scenarios.
For high-accuracy, ARGMAX outperformed ALL.
ALL slightly underperformed the ONLY\_PRED baseline as well.
This suggests that ARGMAX explanations successfully guided users to follow the prediction of StockAI
while ALL toned down the guidance, which worked negatively in this scenario.
On the other hand, ALL outperformed ARGMAX and the baselines for low-accuracy.
Interestingly, ARGMAX also outperformed the baselines,
which suggests that explanations successfully provide users with insights into situations and AI accuracy and can contribute to better decision-making.
ALL positively worked for low-accuracy by providing multiple perspectives.

\subsection{Experiment with {\PropMethod}}
\subsubsection{Procedure}
\begin{figure}[t]
  \begin{center}
    \begin{minipage}[b]{0.75\linewidth}
      \centering
      \includegraphics[width=0.99\linewidth]{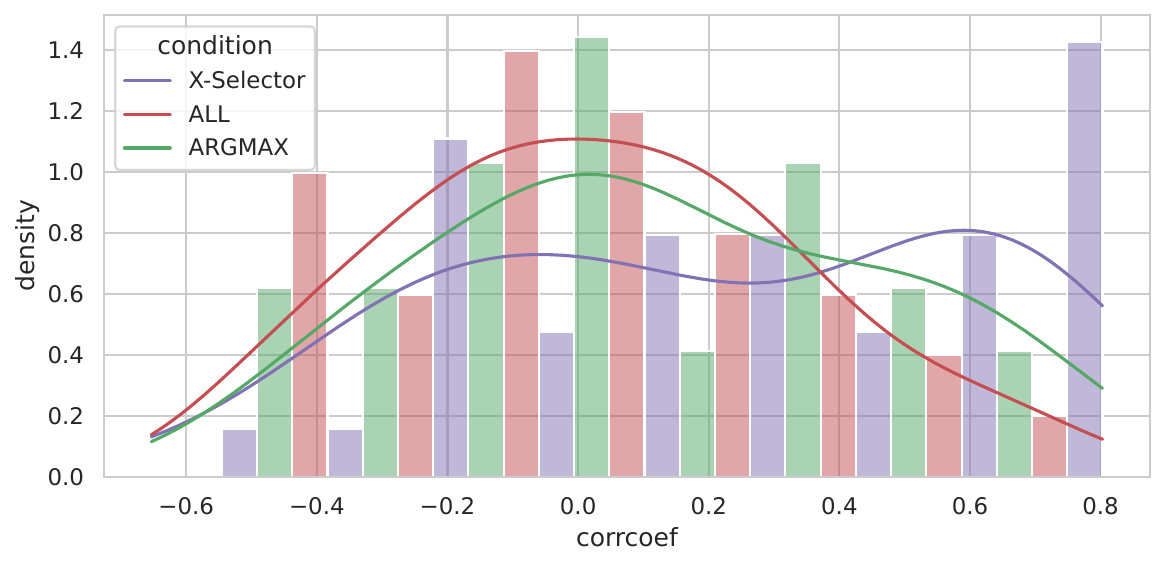}
      \subcaption{Distribution of correlation coefficient between $d_u$ and $d_\mathrm{AI}$ for each user}
      \label{fig:corrcoef-1516}
    \end{minipage}
    \begin{minipage}[b]{0.85\linewidth}
      \centering
      \includegraphics[width=0.99\linewidth]{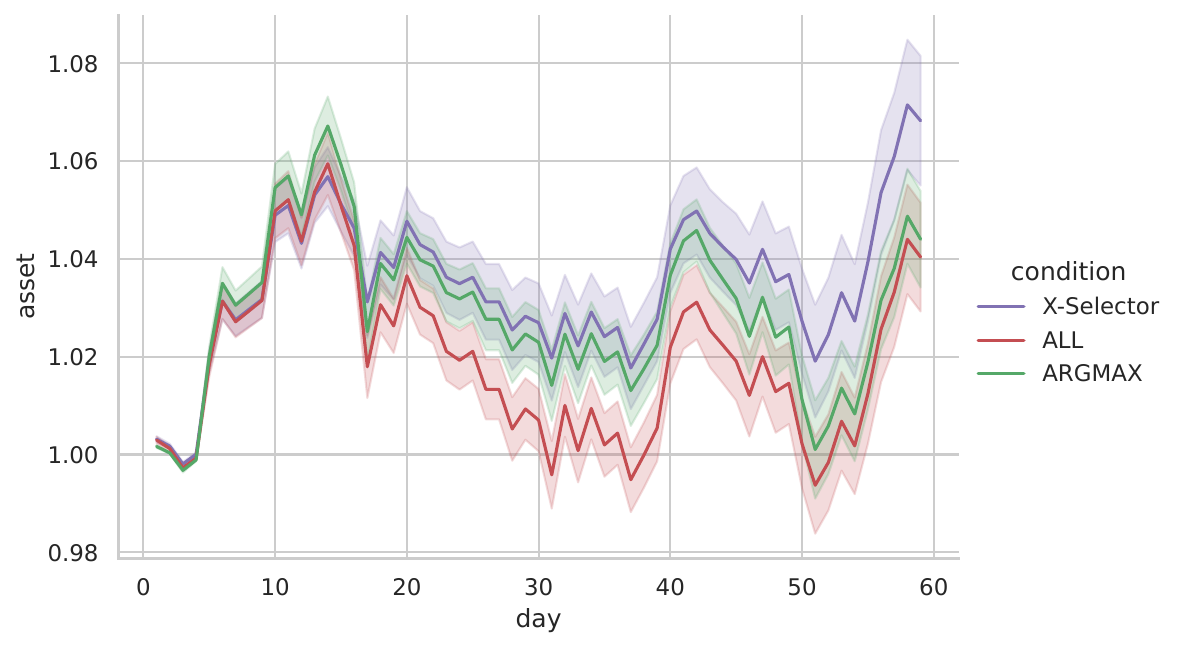}
      \subcaption{User performance}
      \label{fig:pilot-1516}
    \end{minipage}
    \caption{Results for high-accuracy scenario}\label{fig:high}
      \vspace{-12pt}
  \end{center}
\end{figure}
To evaluate {\PropMethod}, we conducted a simulation with its selected explanations. 
To train UserModel, we used the data of the preliminary experiment and additional data acquired in another experiment in which explanations were randomly selected.
We added the data to broaden the variety of explanation combinations in the dataset.
The numbers of the additional participants were 54 and 45 for high- and low-accuracy, respectively.
We conducted a 4-fold cross validation for UserModel, and the correlation coefficient between the model's predictions and the ground truths was 0.429 on average ($SD = 0.056$).

We obtained a participation of 97 participants. Finally, 39 and 35 participants completed the task for high-accuracy and low-accuracy, respectively (46 males, 26 females and 2 did not answer; aged 23-64, $M = 39.6, SD = 10.0$). 

To analyze the results, we compared the correlation coefficient between $d_\mathrm{AI}$ and $d_u$ for each participant as a measure of whether {\PropMethod} could successfully guided users to $d_\mathrm{AI}$,
as well as the comparison of user performance that we did in the preliminary experiment.

\subsubsection{Result}
Figure \ref{fig:corrcoef-1516} shows the correlation coefficient distribution between $d_\mathrm{AI}$ and $d_u$ in the high-accuracy condition. 
The results for ALL and ARGMAX are also shown for comparison.
Notably, while the peaks for ALL and ARGMAX are centered around zero, {\PropMethod} shifted this peak to the right, indicating a stronger correlation between $d_u$ and $d_\mathrm{AI}$ for a greater number of participants.
This means that {\PropMethod} effectively guides users to $d_\mathrm{AI}$ without coercing but presenting explanations selectively.

Figure \ref{fig:pilot-1516} illustrates the user performance.
{\PropMethod} generally outperformed ALL and ARGMAX, meaning that {\PropMethod} enabled users to trade better with selective explanations.
In more detail, {\PropMethod} first underperformed ARGMAX, but the score reversed on day 16.
The gap once narrowed near day 39, but it broadened again until the end.

A possible reason for {\PropMethod}'s better performance is that it can predict which combination of explanations guides participants to sell or buy shares more.
For example, the stock price around day 16 dropped steeply, so the IDSS needed to guide participants to reverse their position.
Here, whereas ARGMAX showed explanations for BEAR, {\PropMethod} showed explanations for NEUTRAL as well as BEAR, which may have helped users sell their shares more. 
Similarly, {\PropMethod} also attempted to guide users to buy a moderate amount when $d_{\mathrm{AI}}$ was positive but not high by, for example, showing only a saliency map for BULL and no text explanations.
Another reason is that {\PropMethod} can overcome the ambiguity in the interpretation of $p$.
$p$ reflects a momentum of stock price in the high-accuracy scenario and must provide some insight for trading,
but it was up to the participants how to use this to actually decide their order.
$\pi$ sometimes suggested that they buy shares even though NEUTRAL or BEAR was the most likely in $p$.
Thus, we can say that $p$ was poorly calibrated, but by referring to $\pi$, {\PropMethod} can avoid misleading participants and instead lead them to more promising decisions.

On the other hand, {\PropMethod} underperformed ARGMAX until day 16.
The stock price was in an uptrend until day 14, and ARGMAX continuously presented explanations for BULL for 12 days in a row,
which may have strongly guided participants to buy stocks and lead to large benefits.
In our implementation, UserModel considers only the explanations of the day and does not consider the history of what explanations were previously provided,
which can be a next target for future work.
\begin{figure}[t]
  \begin{center}
    \begin{minipage}[b]{0.75\linewidth}
      \centering
      \includegraphics[width=0.99\linewidth]{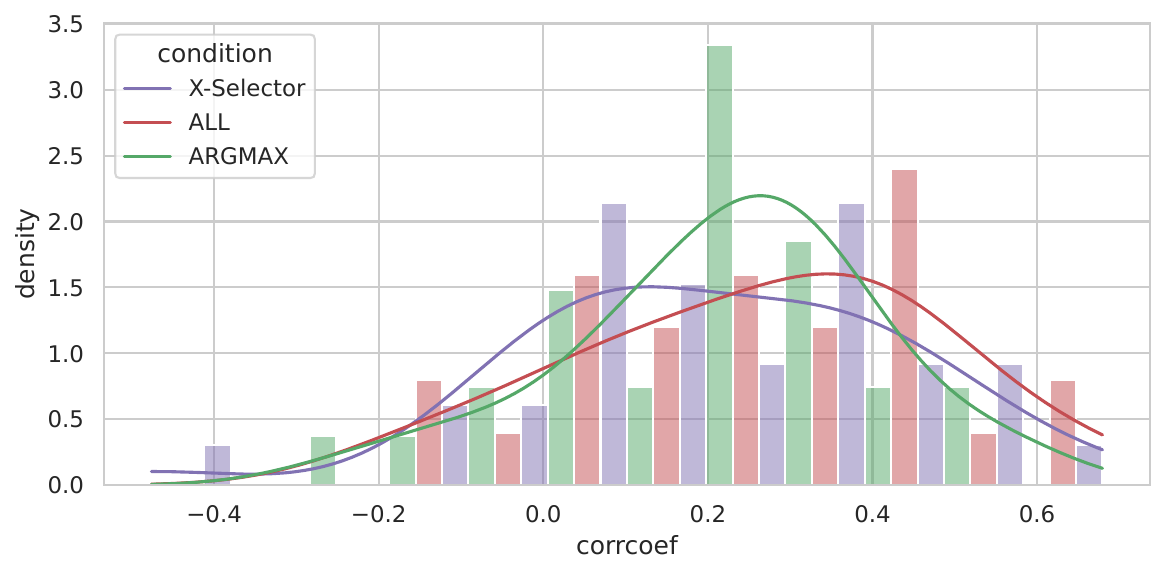}
      \subcaption{Distribution of correlation coefficient}
      \label{fig:corrcoef-2030}
    \end{minipage}
    \begin{minipage}[b]{0.85\linewidth}
      \centering
      \includegraphics[width=0.99\linewidth]{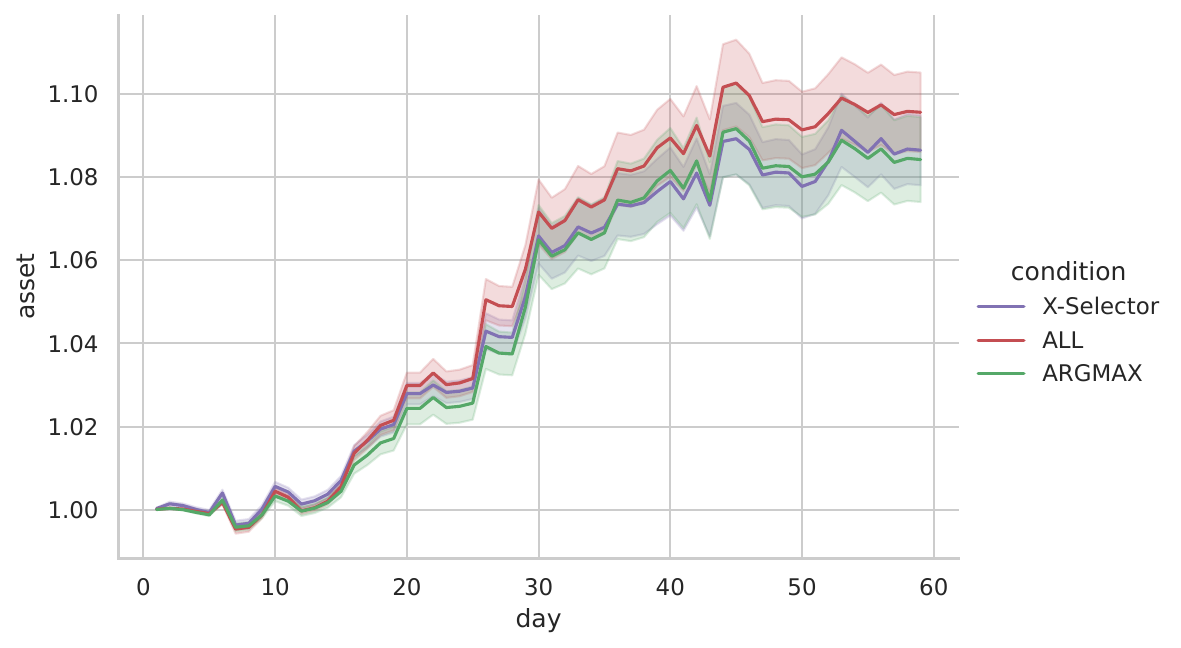}
      \subcaption{User performance}
      \label{fig:pilot-2030}
    \end{minipage}
    \caption{Results for low-accuracy scenario}\label{fig:low}
      \vspace{-12pt}
  \end{center}
\end{figure}

{\PropMethod} could not improve user performance in the low-accuracy scenario (Figure~\ref{fig:pilot-2030}).
Overall, the result was similar to ARGMAX and underperformed ALL.
We further focused on the correlation coefficient between $d_u$ and $d_\mathrm{AI}$.
Figure~\ref{fig:corrcoef-2030} shows that, contrary to the high-accuracy scenario, {\PropMethod} did not increase the score.
The different results between the high- and low-accuracy scenarios indicate the possibility that participants actively assessed the reliability of the AI and autonomously decided whether to follow {\PropMethod}'s guidance.
This itself highlights a positive aspect of introducing libertarian paternalism to human-AI interaction in that users can potentially avoid AI failure depending on its reliability.
However, this did not result in improving their performance in this scenario.
The lack of correlation between the score and the final asset amounts in the {\PropMethod} condition ($r = 0.048$) suggests that merely disregarding the AI's guidance does not guarantee performance improvement.
A future direction for this problem can be developing a mechanism to control the strength of AI guidance and provide explanations in more neutral way depending on AI accuracy.

\section{Conclusion}
This paper investigated the question of how IDSSs can select explanations, and we proposed {\PropMethod},
which is a method for dynamically selecting which explanations to provide along with an AI prediction.
In {\PropMethod}, UserModel predicts the effect of presenting explanations on a user decision for each possible combination to show.
Then, it selects the best combination that minimizes the difference between the predicted user decision and the AI's suggestion.
We applied {\PropMethod} to a stock trading simulation with the support of an XAI-based IDSS.
The result indicated that {\PropMethod} can select explanations that guide users to suggested decisions effectively and improve the performance when the accuracy of the AI is high,
and in addition, it revealed a new challenge of {\PropMethod} for low-accuracy cases.

\bibliography{base}
\bibliographystyle{IEEEtran}

\end{document}